\documentclass[conference]{IEEEtran}
\IEEEoverridecommandlockouts
\usepackage{cite}
\usepackage{amsmath,amssymb,amsfonts}
\usepackage{algorithmic}
\usepackage{graphicx}
\usepackage{textcomp}
\usepackage{xcolor}
\usepackage{url}            
\usepackage{hyperref}

\def\BibTeX{{\rm B\kern-.05em{\sc i\kern-.025em b}\kern-.08em
    T\kern-.1667em\lower.7ex\hbox{E}\kern-.125emX}}

\begin{document}

\title{COVID-MobileXpert: On-Device COVID-19 Patient Triage  and Follow-up using Chest X-rays}

\author{\IEEEauthorblockN{Xin Li}
\IEEEauthorblockA{\textit{Department of Computer Science} \\
\textit{Wayne State University}\\
Detroit, MI 48202 \\
Email: \href{mailto:xinlee@wayne.edu}{xinlee@wayne.edu}}
\and
\IEEEauthorblockN{Chengyin Li}
\IEEEauthorblockA{\textit{Department of Computer Science} \\
\textit{Wayne State University}\\
Detroit, MI 48202 \\
Email: \href{mailto:cyli@wayne.edu}{cyli@wayne.edu}}
\and
\IEEEauthorblockN{Dongxiao Zhu}
\IEEEauthorblockA{\textit{Department of Computer Science} \\
\textit{Wayne State University}\\
Detroit, MI 48202 \\
Email: \href{mailto:dzhu@wayne.edu}{dzhu@wayne.edu}}

}
\maketitle

\begin{abstract}

During the COVID-19 pandemic, there has been an emerging need for rapid, dedicated, and point-of-care COVID-19 patient disposition techniques to optimize resource utilization and clinical workflow. In view of this need, we present COVID-MobileXpert: a lightweight deep neural network (DNN) based mobile app that can use chest X-ray (CXR) for COVID-19 case screening and radiological trajectory prediction. We design and implement a novel three-player knowledge transfer and distillation (KTD) framework including a pre-trained attending physician (AP) network that extracts CXR imaging features from a large scale of lung disease CXR images, a fine-tuned resident fellow (RF) network that learns the essential CXR imaging features to discriminate COVID-19 from pneumonia and/or normal cases with a small amount of COVID-19 cases, and a trained lightweight medical student (MS) network to perform on-device COVID-19 patient triage and follow-up. To tackle the challenge of vastly similar and dominant fore- and background in medical images, we employ novel loss functions and training schemes for the MS network to learn the robust features. We demonstrate the significant potential of COVID-MobileXpert for rapid deployment via extensive experiments with diverse MS architecture and tuning parameter settings. The source codes for cloud and mobile based models are available from the following url:\url{https://github.com/xinli0928/COVID-Xray}.

\end{abstract}

\begin{IEEEkeywords}
COVID-19, SARS-CoV-2, On-device Machine Learning, Trajectory Prediction, Chest X-Ray (CXR)
\end{IEEEkeywords}

\section{Introduction}
\label{sec:introduction}

Due to its flu-like symptoms and potentially serious outcomes, a dramatic increase of suspected COVID-19 cases are expected to overwhelm the healthcare system during the flu season. Health systems still largely allocate facilities and resources such as Emergency Department (ED) and Intensive Care Unit (ICU) on a reactive manner facing significant labor and economic restrictions. To optimize resource utilization and clinical workflow, a rapid, automated, and point-of-care COVID-19 patient management technology that can triage (COVID-19 case screening) and follow up (radiological trajectory prediction) patients is urgently needed.

Chest X-ray (CXR), though less accurate than a PCR diagnostic, chest Computed Tomography (CT) or serological test, became an attractive option for patient management due to its impressive portability, availability and scalability \cite{wong2020frequency}. At present, the bottleneck lies in the shortage of board certified radiologists who are capable of identifying massive COVID-19 positive cases to reduce wait time at ED and determining the radiological trajectory of the COVID-19 patients after admission. The intensive development of deep neural network (DNN) powered CXR image analysis has seen the unprecedented success in automatic classification and segmentation of lung diseases \cite{wang2017chestx}. Using the cloud solutions such as Google Cloud Platform or on-premise computing clusters to train a sophisticated DNN (e.g., DenseNet-121 \cite{huang2017densenet}) with dozens of millions of parameters and hundreds of layers via billions of operations for both training and inference, these large scale Artificial Intelligence (AI) models achieve amazing performance that even outperforms board certified radiologists in some well-defined tasks \cite{rajpurkar2017chexnet}.  

With the increasing number of smart devices and improved hardware, there is a growing interest to deploy machine learning models on the device to minimize latency and maximize the protection of privacy. However, up to date on-device medical imaging applications are very limited to basic functions, such as the DICOM image view. In the COVID-19 environment, a mobile AI approach is expected not only to protect patient privacy, but also to provide a rapid, effective and efficient assessment of COVID-19 patients without the immediate need for an on-site radiologist. However, a major challenge that prevents wide adoption of the mobile AI approach is lack of lightweight yet accurate and robust neural networks.

Adequate knowledge has been accumulated from training the large scale DNN systems to accurately discern the subtle difference among the different lung diseases by learning the discriminative CXR imaging features \cite{rajpurkar2017chexnet}. Leveraging these results, we design and implement a novel three-player knowledge transfer and distillation (KTD) framework composed of an Attending Physician (AP) network, a Resident Fellow (RF) network, and a Medical Student (MS) network for on-device COVID-19 patient triage and follow-up. In a nutshell, we pre-train a full AP network using a large scale of lung disease CXR images \cite{wang2017chestx,rajpurkar2017chexnet}, followed by fine-tuning a RF network via knowledge transfer using labeled COVID-19, pneumonia and normal CXR images, then we train a lightweight MS network for on-device COVID-19 patient triage and follow-up via knowledge distillation. After the KTD framework, the lightweight MS network is able to produce expressive features to identify COVID-19 cases as well as predict the radiological trajectory. The unique features of the KTD framework are knowledge transfer from large-scale existing lung disease images to enhance expressiveness of learned representation and novel loss functions to increase robustness of knowledge distillation to the MS network. 

To the best of our knowledge, currently, there is no mobile AI system for on-device COVID-19 patient triage and follow-up using CXR images. In this work, we present an AI-powered system, COVID-MobileXpert, to triage and follow up COVID-19 patients using portable X-rays at the patient’s location. At the ED, COVID-MobileXpert calculates COVID-19 probabilistic risk to assist automated triage of COVID-19 patients. At the ICU or general ward (GW), it uses a series of longitudinal CXR images to determine whether there is an impending deterioration in the health condition of COVID-19 patients. Therefore, COVID-MobileXpert is essential to fully realize the potential of CXR to exert both immediate and long-term positive impacts on US healthcare systems. It enjoys the following advantages: 1) accurately detecting positive COVID-19 cases particularly from closely related pneumonia cases; 2) continuously following up admitted patients via radiological trajectory prediction.

\section{Related Work}
\begin{figure}
	\begin{center}
		\includegraphics[width=0.48\textwidth]{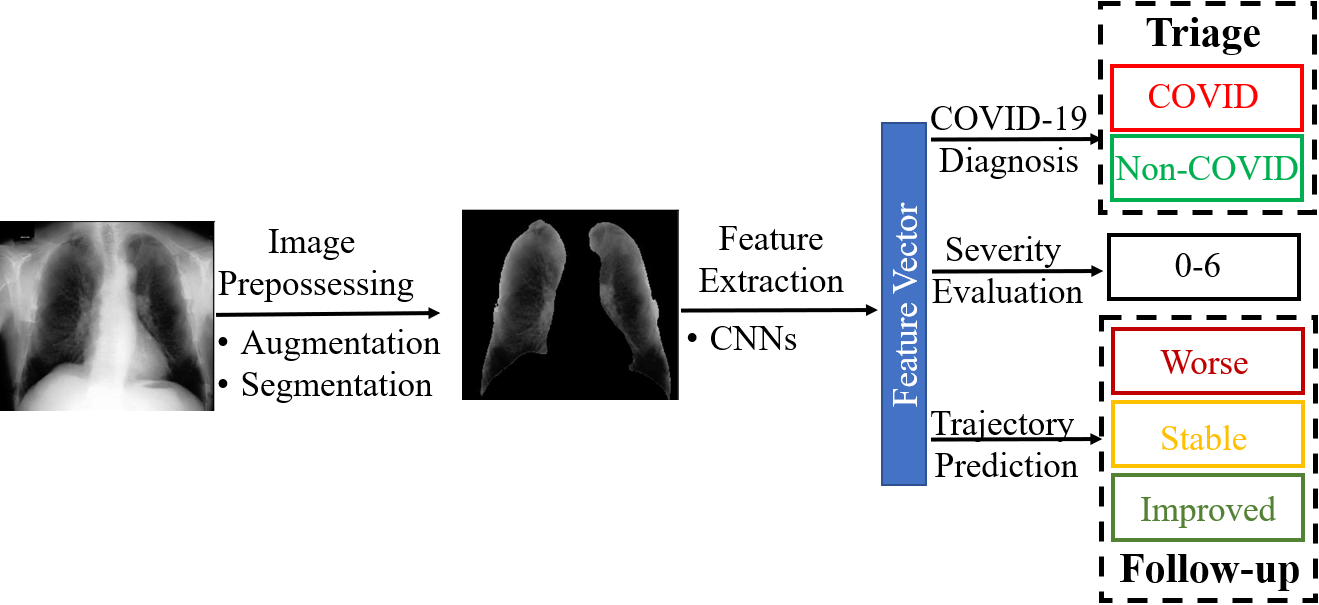}
	\end{center}
	\caption{\small{The analytical workflow of COVID-19 CXR interpretation systems.}}\label{fig:reivew}\vspace{-5mm}
\end{figure}

\subsection{COVID-19 CXR Interpretation}

During the COVID-19 pandemic in the past few months, convolutional neural networks (CNNs) have been successfully employed to assist with COVID-19 CXR interpretation. Since mid-February 2020, we collected over 40 works (most of them in a pre-print format) focusing on this subject. Although they may focus on different specific tasks, as shown in Fig. \ref{fig:reivew}, they follow a similar pipeline: image prepossessing, feature extraction, and interpretation. 

Data augmentation and segmentation are widely used as part of these approaches to avoid overfitting. In addition to the basic transformation-based method which includes rotating, flipping, scaling used in \cite{chowdhury2020can}, Khalifa et al. \cite{khalifa2020detection} applied Generative Adversarial Network to generate virtual samples for data augmentation. To reduce the interference caused by unrelated area, Yeh et al.\cite{yeh2020cascaded}, Lv et al.\cite{lv2020cascade} and Signoroni et al. \cite{signoroni2020end} applied a U-net based method to perform a fast lung segmentation and preserved the region of interest (ROI) only. After image prepossessing, discriminative patterns for COVID-19 such as ground-glass opacification/opacity (GGO) are then extracted by CNNs. Most current studies have directly borrowed or adopted well-known architectures such as ResNet \cite{chowdhury2020can,hall2020finding}, InceptionV3 \cite{chowdhury2020can}, DenseNet \cite{yeh2020cascaded,lv2020cascade,cohen2020predicting}, and VGG \cite{hall2020finding,zhu2020deep}.

After feature extraction, three major tasks have been performed: diagnosis, severity evaluation, and trajectory prediction. COVID-19 diagnosis is usually considered to be a classification problem. The most straightforward way of detecting COVID-19 is to train a classifier with cross-entropy loss which is applied within most of these approaches. Other than these straightforward approaches, Zhang et al. \cite{zhang2020viral} employed an unsupervised anomaly detection approach that detects COVID-19 cases as outliers. Moreover, Hassanien et al. \cite{hassanien2020automatic} developed a classifier based on a support vector machine. 

As for severity evaluation, Cohen et al. \cite{cohen2020predicting} and Zhu et al. \cite{zhu2020deep} directly predicted lung disease severity scores using a linear regression model based on extracted features. In order to associate each score with a confidence value, Signoroni et al. \cite{signoroni2020end} treated this task as a joint multi-class classification and regression problem using a compound loss function. Based on the severity assessment, the trajectory prediction can be achieved by calculating the difference in severity score between two adjacent CXR images. Other than basic score level interpretation, Duchesne et al. \cite{duchesne2020tracking} built their trajectory prediction model based on feature level. When the feature from a single CXR was extracted by DenseNet-121, they used logistic regression to classify the trajectory into one of three categories: ``Worse", ``Stable", or ``Improved". However, the feature from a single CXR may not be sufficient to predict radiological trajectory. In order to tackle these challenges, our model forecasts radiological trajectory using feature extracted from a series of longitudinal CXR images of a single patient. By incorporating longitudinal CXR images into our model, novel imaging features of progressive disease, including subtle changes of radiological features that are invisible to the human eye, can be detected.
\begin{figure*}
	\begin{center}
		\includegraphics[width=1\textwidth]{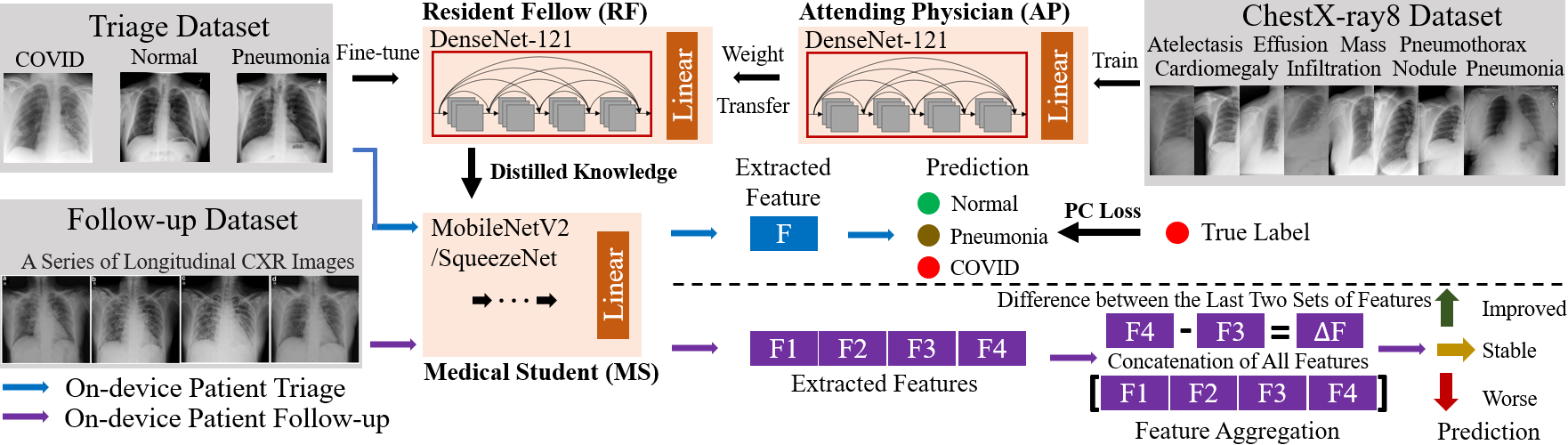}
	\end{center}
	\caption{\small{Overview of the three-player KTD training architecture demonstrating the {\bf knowledge transfer} from AP to RF and the {\bf knowledge distillation} from RF to MS. The blue and purple arrows demonstrate the training for two tasks: patient triage and follow-up respectively. }}\label{fig:transfer}	\vspace{-3mm}

\end{figure*}
\subsection{On-device AI Model}

Currently, most AI models trained for COVID-19 interpretation are full DNNs that are not suitable to deploy on resource-constrained mobile devices. As there is no existing on-device medical image interpretation research, the vast majority of the existing work focuses on comparing the performance of different lightweight neural networks such as MobileNetV2 \cite{sandler2018mobilenetv2}, SquezzeNet \cite{iandola2016squeezenet}, Condense-Net \cite{huang2018condensenet}, ShuffleNetV2 \cite{zhang2018shufflenet}, MnasNet \cite{tan2019mnasnet} and MobileNetV3 \cite{howard2019searching} using small benchmark natural image datasets such as CIFAR 10/100. MnasNet and MobileNetV3 are representative models generated via automatic neural architecture search (NAS) whereas all other networks are manually designed \cite{elsken2018neural}. Due to the practical hardware resource constraint of mobile devices, natural image classification and segmentation performance have been compared based on accuracy, energy consumption, runtime, and memory complexity that no single network has demonstrated superior performance in all tasks \cite{dhar2019device}. Besides tailor-made network architectures for mobile devices, compression of the full DNN at the different stages of training also stands as a promising alternative. For in-training model compression, for example, Chen et al. \cite{chen2019drop} designed a novel convolution operation via factorizing the mixed feature maps by their frequencies to store and process feature maps that vary spatially slower at a lower spatial resolution to reduce both memory and computation cost of the image classification. Post-training or fine-tuning model compression techniques such as quantization \cite{rastegari2016xnor} and/or pruning techniques \cite{han2015deep} are often used to reduce the model size at the expense of reduced prediction accuracy. Wang et al. \cite{wang2018training} demonstrated using 8-bit floating-point numbers for representing weight parameters without compromising the model's accuracy. Lou et al. \cite{lou2020auto} automatically searched a suitable precision for each weight kernel and chose another precision for each activation layer and demonstrated a reduced inference latency and energy consumption while achieving the same inference accuracy. Tung and Mori \cite{tung2018clip} combined network pruning and weight quantization in a single learning framework to compress several DNNs without sacrificing accuracy.  

In order to improve the performance of the lightweight on-device models, knowledge distillation \cite{hinton2015distilling} is also used where a full teacher model is trained on the cloud or an on-premise GPU cluster, and a student model is trained at the mobile device with the `knowledge' distilled via the soft labels from the teacher model. Thus the student model is trained to mimic the outputs of the teacher model as well as to minimize the cross-entropy loss between the true labels and predictive probabilities (soft labels). Knowledge distillation yields compact student models that outperform the compact models trained from scratch without a teacher model \cite{phuong2019towards}. Goldblum et al. \cite{goldblum2019adversarially} attempted to encourage the student network to output correct labels using the training cases crafted with a moderate adversarial attack budget to demonstrate the robustness of knowledge distillation methods. Unlike the natural images, on-device classification of medical images remain largely an uncharted territory due to the following unique challenges: 1) label scarcity in medical images significantly limits the generalizability of the machine learning system; 2) vastly similar and dominant fore- and background in medical images make it hard samples for learning the discriminating features between different disease classes. To tackle these unique challenges we propose a novel three-player framework for training a lightweight network towards accurate and hardware friendly on-device COVID-19 patient triage and follow-up.

\section{Method}
\label{sec:model}

\subsection{Model Architecture}
We employ DenseNet-121 architecture as the template to pre-train and fine-tune the AP and RF networks. In addition, among well-studied lightweight CNNs \cite{dhar2019device}, we select the most well-applied network MobileNetV2, and the most lightweighted network SqueezeNet as the candidate MS networks for on-device COVID-19 case screening and radiological trajectory prediction. Table \ref{tab:hardware} summarizes the key model complexity parameters \cite{dhar2019device}. Fig. \ref{fig:transfer} illustrates the three-player KTD training framework where the knowledge of abnormal CXR images is transferred from AP network to RF network and knowledge of discriminating COVID-19, non-COVID-19, and pneumonia is distilled from the RF network to the MS network. 

\begin{table}[]
	\begin{center}
		\begin{tabular}{lllll}
			\hline
			Metric               &DenseNet-121     & MobileNetV2  & SqueezeNet  \\\hline
			\# of CONV layers 	& 120	  & 20                   & 22         \\
			Total weights     &  7.9M      & 3.47M             & 0.72M      \\
			Total MACs          &   2900M   & 300M              & 282M      \\
			\hline
		\end{tabular}
	\end{center}
	\caption{Comparison of full and compact DNN model complexity.}
	\label{tab:hardware}\vspace{-8mm}
\end{table}

\subsection{The KTD Training Scheme}
We pre-train the AP network as the {\it source task}, i.e., lung disease classification, and fine-tune, validate and test the RF network as the {\it destination task}. Different from recent studies \cite{wang2020covid} that pre-train the models with natural image datasets such as ImageNet, we pre-train the DenseNet-121 based AP network using the more related ChestX-ray8 dataset \cite{wang2017chestx} of 108,948 lung disease cases to extract the CXR imaging features of lung diseases instead of generic natural imaging features. Specifically, beyond the dense block, we employ a shared fully connected layer for extracting the general CXR imaging feature and 8 fully connected disease-specific layers (including pneumonia as one disease layer) to extract disease-specific features (Fig. \ref{fig:transfer}). Following the pre-training using large ChestX-ray8 dataset, the weights defining the general CXR imaging feature and the pneumonia disease feature are transferred to fine-tune the DenseNet-121 based RF network using a smaller compiled dataset of 3 classes of CXR images, i.e., COVID-19, normal and pneumonia. The latter is randomly initialized using two sets of weight parameters corresponding to normal and COVID-19 classes with the initial values of other weight parameters transferred from the pre-trained source model. The RF network is then used to train the lightweight MS network, e.g., MobileNetV2, or SqueezeNet, via knowledge distillation.

As shown in the MS section in Fig. \ref{fig:transfer}, after knowledge distillation, the trained MS network can triage patients by screening COVID-19 cases following the blue arrow. Then a radiological trajectory prediction model is further developed based on the trained MS network. Following the purple arrow, given a series of longitudinal CXR images from one patient, all images are fed into the pre-trained MS network for extracting disease-specific features. These features are then aggregated using different schemes before prediction. Here we investigate two different schemes: 1) calculating the difference between the last two CXR images' features; 2) chronologically concatenating all features. After feature aggregation, two fully connected layers are randomly initialized and trained with softmax loss function for the trajectory prediction.

\subsection{Loss Functions}
As stated above, a unique challenge in medical imaging classification is the so-called ``hard sample problem" \cite{li2020learning}, i.e., a subtle difference on the ROI across the images with a large amount of shared fore- and backgrounds. Motivated by this, we use an in-house developed loss function, i.e., Probabilistically Compact (PC) loss, for training the MS model and compared with ArcFace \cite{deng2019arcface}, the additive angular margin loss for deep face recognition, using the classical softmax loss as the baseline. Both PC and ArcFace losses are designed for improving classification performance on hard samples. PC loss is to encourage the maximized margin between the most probable label (predictive probability) and the first several most probable labels whereas ArcFac loss is to encourage  widening the geodesic distance gap between the closest labels. In terms of predicted probabilities, DNN robustness is beneficial from the large gap between $f_y(\boldsymbol{x})$ and $f_k(\boldsymbol{x})$ $(k\neq y)$, where $f_y(\boldsymbol{x})$ represents the true class and $f_k(\boldsymbol{x})$ $(k\neq y)$ represents the most probable class. Indeed, the theoretical study \cite{neyshabur2017exploring} in deep learning shows that the gap $f_y(\boldsymbol{x})-\max_k f_k(\boldsymbol{x})$ can be used to measure the generalizability of DNNs.

The PC loss to improve CNN's robustness is as follows:
\begin{equation}\label{eq:pcl}
L_{pc}(\theta) =\frac{1}{N}\sum_{k=1}^{K}\sum_{i_k\in S_k} \sum_{j=1,j\neq k}^{K}\max\{0, f_j(\boldsymbol{x}_{i_k}) +\xi -f_k(\boldsymbol{x}_{i_k})\},
\end{equation}
where $N$ is the number of training samples, $\xi >0$ is the probability margin treated as a hyperparameter. Here, we include all non-target classes in the formulation and penalize any classes for each training sample that violate the margin requirement for two reasons: (1) by maintaining the margin requirement for all classes, it provides us convenience in implementation as the first several most probable classes can change during the training process; and (2) if one of the most probable classes satisfies the margin requirement, all less probable classes will automatically satisfy this requirement and hence have no effect on the PC loss. Compared with previous works that explicitly learn features with large inter-class separability and intra-class compactness, the PC loss avoids assumptions on the feature space, instead, it only encourages the feature learning that leads to probabilistic intra-class compactness by imposing a probability margin $\xi$. 

\section{Experiment and Results}
\label{sec:experiments}
In Section \ref{sec:experiments}, we design and conduct extensive experiments to evaluate the performance of the compact MS network in patient triage and follow-up. In order to gain a holistic view of the model behavior, we investigate the performance concerning multiple choices of loss functions and values of tuning parameters for COVID-19 case screening as well as various choices of feature aggregation schemes and classifiers for radiological trajectory prediction. 
\begin{figure}
	\begin{center}
		\includegraphics[width=0.4\textwidth]{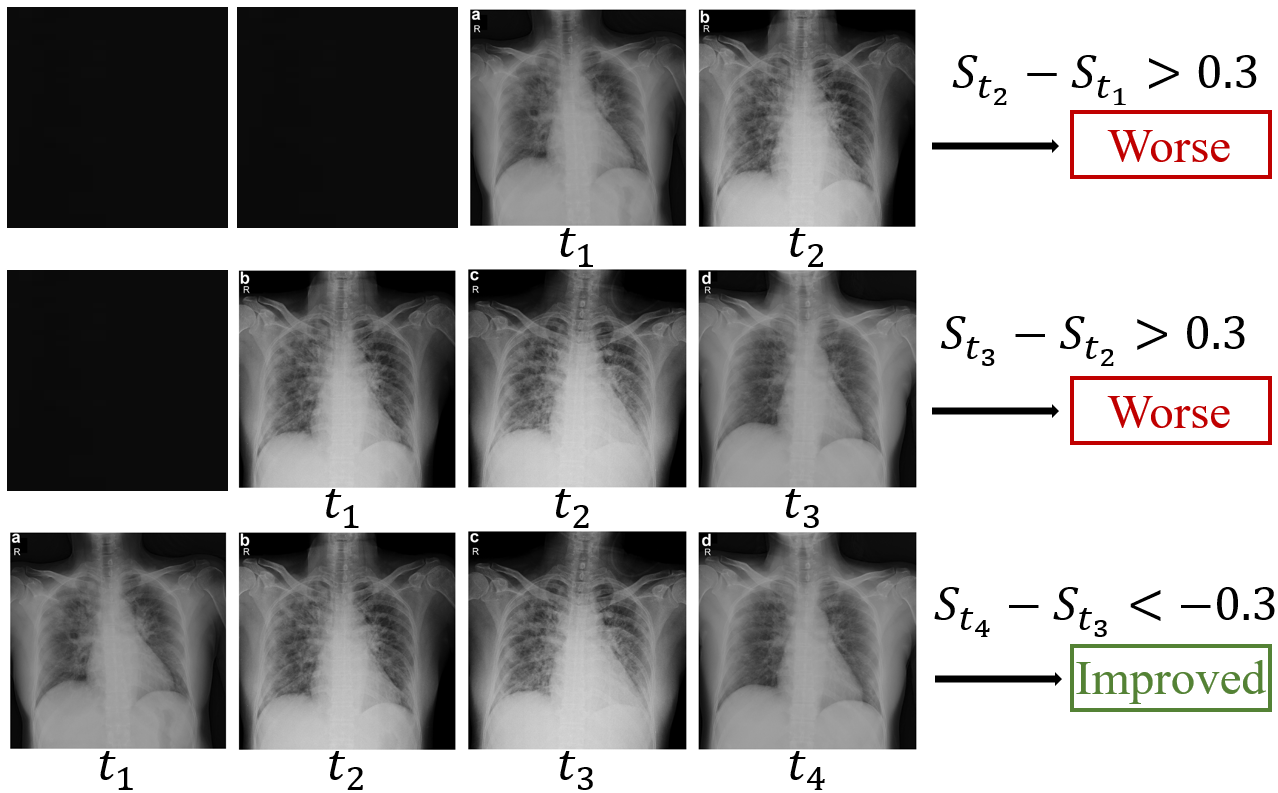}
	\end{center}
	\caption{\small{An example of data preparation for a series of longitudinal CXR images with radiological trajectory labels. The patient is in critical condition on $t_3$ then recovered afterward.}}\label{fig:label}
		\vspace{-3mm}

\end{figure}
\subsection{Datasets}
The CXR image dataset for COVID-19 patient triage is composed of 179 CXR images from normal class \cite{kaggle_2018}, 179 from pneumonia class \cite{kaggle_2018} and 179 from COVID-19 class containing both PA (posterior anterior) and AP (anterior posterior) positions \cite{cohen2020covid} and we split it into training/validation/testing sets with 125/18/36 cases (7:1:2) in each class. Since some patients have multiple CXR images in COVID-19 class, we sample images per patient for each split to avoid images from the same patient being included in both training and test sets.

For the radiological trajectory dataset, we assign a opacity score $S$ for each COVID-19 positive CXR image in \cite{cohen2020covid} using the scoring system provided by \cite{cohen2020predicting}. Fig. \ref{fig:label} shows an example of how we generate CXR image sequences and assign corresponding radiological trajectory labels (i.e., ``Worse", ``Stable", ``Improved"). Given a COVID-19 patient's CXR images over four time points (the maximum length is set to four time points), we can create three CXR image sequences with zero-padding. For each sequence, we calculate the difference in the score of the last two CXR images. If the difference is larger than $0.3$ the sequence is categorized as ``Worse", if the difference is less than $-0.3$, it is labeled as ``Improved", otherwise, the category is ``Stable". We collect a total of 159 CXR image sequences from 100 patients in \cite{cohen2020covid} and the dataset contains 80 ``Worse" samples, 28 ``Stable" samples, and 51 ``Improved" samples. Similarly, we split it into training/validation/testing sets with 111/16/32 samples (7:1:2).

\subsection{Implementation Details}

We implement our model on a GeForce GTX 1080ti GPU platform using PyTorch. The network is trained with the Adam optimizer for 50 epochs with a mini-batch size of 32 (triage task) and 10 (follow-up task). The parameter values that give rise to the best performance on the validation dataset are used for testing. Similar to \cite{duchesne2020tracking}, when training the radiological trajectory prediction model, we employ the pre-trained MS network as a feature extractor (fixed weights). To overcome the overfitting problem, we also apply a dropout regularization with a rate of 0.5.

\subsection{Tunning Parameters}
\noindent $\xi$: in the PC loss formula (Eq. \ref{eq:pcl}), a large value encourages the probabilistic intra-class compactness. \\  
\noindent $\alpha$: in knowledge distillation framework \cite{hinton2015distilling,goldblum2019adversarially} (Eq. \ref{eq:kd}), 

\begin{equation}\label{eq:kd}
\begin{aligned}
\min _{\theta} \mathbb{E}_{(X, y) \sim \mathcal{D}}\left[\alpha t^{2} \operatorname{KL}\left(S_{\theta}^{t}\left(X\right), T^{t}(X)\right)\right.& \\
+(1-\alpha) \ell\left(S_{\theta}^{t}(X), y\right)] &,
\end{aligned}
\end{equation}

it regularizes the `strength' of knowledge distillation by specifying the relative contributions of the distillation loss, i.e., $\operatorname{KL}\left(S_{\theta}^{t}\left(X\right), T^{t}(X)\right)$, measuring how well the MS model mimics the RF model's behavior using KL divergence and the classification loss of the MS model, i.e., $\ell\left(S_{\theta}^{t}(X), y\right)$. $S_{\theta}(.)$ and $T(.)$ represent the RF model and MS model, respectively. The larger value, the stronger knowledge distillation is enforced from the RF model to the MS model.   

\noindent $T$: in Eq. \ref{eq:kd}, it represents temperature where $T=1$ corresponds to the standard softmax loss. As the value of $T$ increases, the probability distribution generated by the softmax loss becomes softer, providing more information regarding which classes the RF model found more similar to the predicted class.   
\begin{table}[]
	\begin{center}
		\begin{tabular}{ccccc}
			\hline
			\multicolumn{5}{c}{MobileNetV2/SqueezeNet (T=5)}                                           \\ \hline
			\multicolumn{1}{c|}{$\alpha$} & PC($\xi$ = 0.8)        & PC($\xi$ = 0.995)  & ArcFace  & SM   \\ \hline
			\multicolumn{1}{c|}{0.2}      & 0.870/0.798          & 0.833/0.777      & 0.870/0.750  & 0.861/0.777 \\
			\multicolumn{1}{c|}{0.4}      & \textbf{0.880}/0.777            & 0.870/0.815      & 0.861/0.796   & 0.833/0.759  \\
			\multicolumn{1}{c|}{0.6}      & 0.851/0.796           & 0.851/0.787      & 0.851/0.805   & 0.861/0.796\\
			\multicolumn{1}{c|}{0.8}      & \textbf{0.880}/0.824 & 0.870/0.796      & 0.851/0.796   & 0.833/0.787 \\ \hline
			\multicolumn{5}{c}{MobileNetV2/SqueezeNet ($\alpha=0.8$)}                      \\ \hline
			\multicolumn{1}{c|}{T}  & PC($\xi$ = 0.8)  & PC($\xi$ = 0.995)  & ArcFace  & SM  \\ \hline
			\multicolumn{1}{c|}{1}  & 0.851/0.750   & \textbf{0.880}/0.814      & 0.870/0.796   & 0.870/0.796 \\
			\multicolumn{1}{c|}{5}  & \textbf{0.880}/0.824    & 0.870/0.796      & 0.851/0.796   & 0.833/0.787 \\
			\multicolumn{1}{c|}{10} & \textbf{0.880}/0.796     & 0.842/0.750      & 0.861/0.787   & 0.870/0.824 \\ \hline
		\end{tabular}
	\end{center}
	\caption{Classification performance of MS networks, The values in ./. indicate MobileNetV2 vs. SqueezeNet.}
	\label{tab:MobileNetV2}\vspace{-8mm}
\end{table}
\subsection{Evaluation of COVID-19 Patient Triage Performance}
We first report the classification accuracy to select the best MS model under different values of hyperparameters, followed by systematic evaluation of the model's discriminating power of COVID-19 from non-COVID pneumonia and normal cases using AUROC values. With the knowledge transfer from the AP network pre-trained with a large set of abnormal lung disease cases, the RF network demonstrates a remarkably high accuracy of 0.935 in the classification of CXR images.

Distilling knowledge from the RF network to the lightweight MS network, we observe an impressive performance that a vast majority of accuracy values are well above 0.850 for CXR image classification. Table \ref{tab:MobileNetV2} shows the classification accuracy results of both MobileNetV2 and SqueezneNet architecture with different loss functions and values of tuning parameters. It is clear that the knowledge distillation is essential to train the lightweight MS network without compromising much accuracy since the MS network alone, without knowledge distillation, achieves a baseline classification accuracy of 0.843 (MobileNetV2) and 0.732 (SqueezeNet), which are lower than those with knowledge distillation shown in Table \ref{tab:MobileNetV2}. 

Looking at Table \ref{tab:MobileNetV2} in more detail, we note that the performance of MobileNetV2 and SqueezeNet are not sensitive to the choice of temperatures (T) and strengths of distillation ($\alpha$), however, it is very sensitive to the choice of loss functions. Overall, the PC loss developed in-house that flattens other probable class predictions perform the best across diverse settings of the tuning parameters, indicating the quality of knowledge distilled from the RF network to the MS network plays a pivotal role in training the lightweight MS network to ensure an accurate on-device COVID-19 patient triage. 

\begin{figure*}
	\begin{center}
		\includegraphics[width=0.875\textwidth]{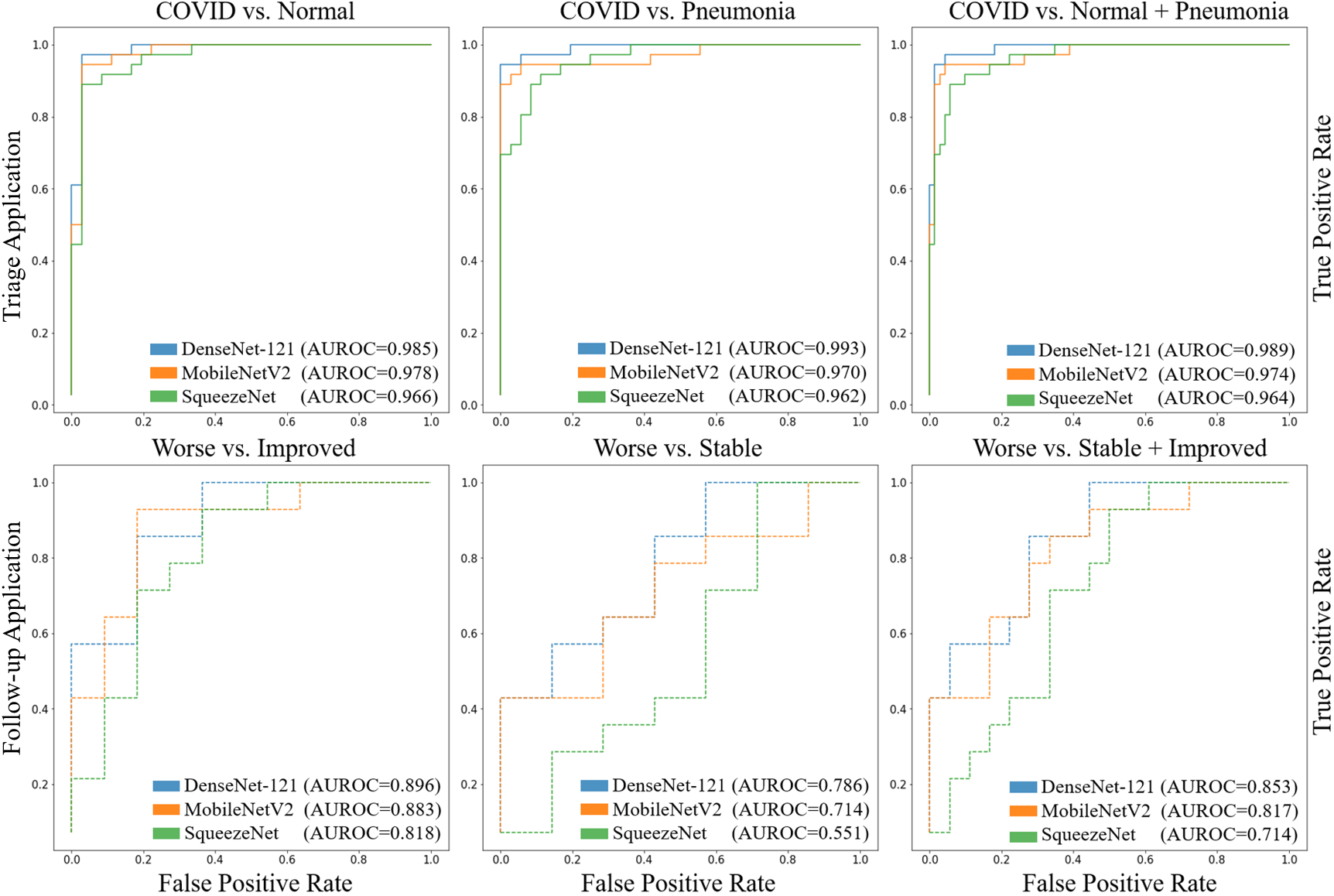}
	\end{center}
	\caption{\small{The upper panel shows the performance of the large-scale RF network and two compact MS networks of discriminating (a) COVID-19 vs. Normal cases; (b) COVID-19 vs. Pneumonia cases and (c) COVID-19 vs. Normal + Pneumonia cases, while the lower panel shows the performance of discriminating (d) ``Worse" vs. ``Improved" cases; (e) ``Worse" vs. ``Stable" cases and (f) ``Worse" vs. ``Improved" + ``Stable" cases. }}\label{fig:ROC}\vspace{-3mm}
\end{figure*}

In order to systematically evaluate the performance of the MS networks under the different decision thresholds, we use the AUROC value to assess how well the model is capable of discriminating COVID-19 cases from normal cases, pneumonia cases as well as normal plus pneumonia cases. In Fig. \ref{fig:ROC}, both compact MS networks, i.e., MobileNetV2 and SqueezeNet, demonstrate a remarkable performance on all discrimination tasks that are comparable to that of the large scale cloud-based RF network, i.e., DenseNet-121. Importantly, both MobileNetV2 and SqueezeNet achieve high AUROC values of 0.970 and 0.964 when discriminating COVID-19 cases against mixed pneumonia and normal cases demonstrating strong potential for on-device triage using CXR images.

\begin{table}[]
\begin{center}
{

\begin{tabular}{ccc}
\hline
\multicolumn{3}{c}{MobileNetV2/SqueezeNet (DenseNet-121)}                    \\ \hline
\multicolumn{1}{c|}{Classifiers} & Difference          & Concatenation       \\ \hline
\multicolumn{1}{c|}{Logistic Regression}          & 0.560/0.640 (0.720) & 0.760/0.720 (0.800) \\
\multicolumn{1}{c|}{Gradient Boosting}          & 0.680/0.640 (0.680) & 0.680/0.680 (0.680) \\
\multicolumn{1}{c|}{Random Forest}          & 0.680/0.600 (0.680) & 0.720/0.680 (0.720) \\
\multicolumn{1}{c|}{Our FC-classifier}        & 0.720/0.680 (0.720) & \textbf{0.800}/0.760 (0.800) \\ \hline
\end{tabular}}
\end{center}
\caption{Performance comparison of two feature aggregation schemes (Difference vs. Concatenation) with four different classifiers using two MS Networks (MobileNetV2 and SqueezeNet) as the feature extractor. Values in parentheses indicate the upper bound of accuracy yielded by RF Network (DenseNet-121).}\label{tab:FL}\vspace{-8mm}
\end{table}	

\subsection{Evaluation of COVID-19 Patient Follow-up Performance}

Similar to \cite{duchesne2020tracking}, we first report the classification accuracy of discriminating ``Worse" versus ``Improved" cases to select the best combination of classifiers and feature aggregation schemes for on-device radiological trajectory prediction, followed by systematic evaluation of the model's discriminating power of ``Worse" cases from ``Improved" and ``Stable“ cases using AUROC values. Based on the features extracted from the MS networks, four classifiers are trained for radiological trajectory prediction: 1) logistic regression; 2) gradient boosting; 3) random forest and 4) MS networks followed by fully connected layers (our FC-classifier).

\begin{figure*}
	\begin{center}
		\includegraphics[width=0.72\textwidth]{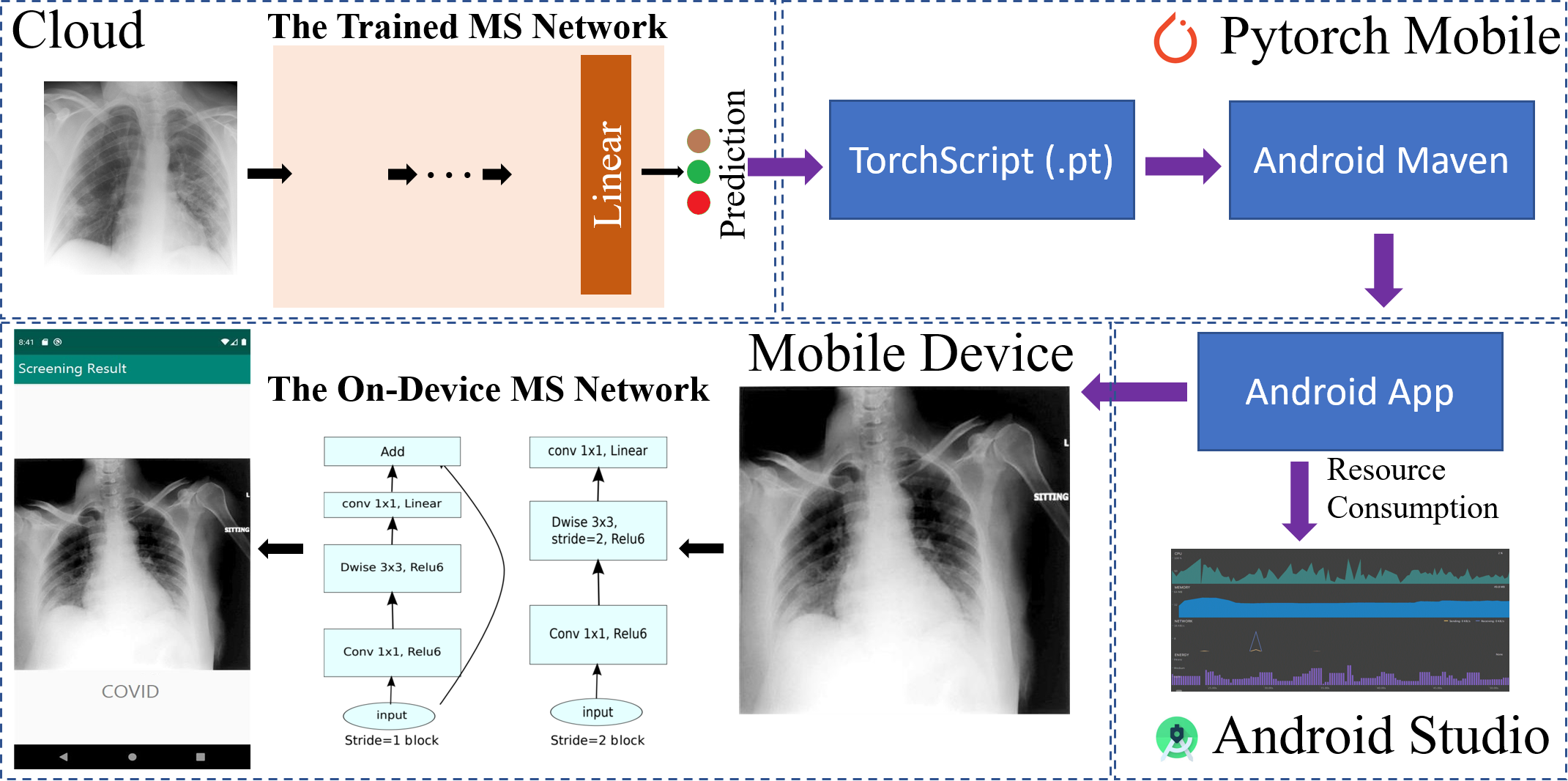}
	\end{center}
	\caption{\small{A schematic overview of on-device deployment of the COVID-MobileXpert.}}\label{fig:android}	

\end{figure*}

As shown in Table \ref{tab:FL}, we observe the classifiers trained based on the feature extracted from both compact MS networks, i.e., MobileNetV2 and SqueezeNet, achieve a very similar level of performance to those trained with the feature from large scale RF network i.e., DenseNet-121. This again demonstrates that KTD training architecture with PC loss performs a high-quality knowledge distillation from RF network to lightweight MS networks. 

When doing comparison between the feature aggregation schemes, we can see a significant improvement from using a series of longitudinal features over using only the difference between the last two sets of features. As for classifier selection, compared with the conventional classifiers i.e., logistic regression, gradient boosting and random forest, our FC-classifier is able to learn a series of subtle changes related to radiological features from CXR images, thus achieving a better performance. As a result, the best on-device performance is obtained by our FC-classifier with feature concatenation using MobileNetV2, which attains the upper bound of accuracy (0.800) yielded by DenseNet-121.
\begin{table*}[]
	\begin{center}
		\scalebox{0.8}{
		\begin{tabular}{c|ccccccccc}
			\hline
			Mobile Systems & \multicolumn{3}{c}{Nexus One} & \multicolumn{3}{c}{Pixel}    & \multicolumn{3}{c}{Pixel 2 XL} \\ \hline
			The MS Network     & CPU (\%)  & Memory (MB) & Energy & CPU (\%) & Memory (MB) & Energy & CPU (\%)  & Memory (MB) & Energy \\ \hline
			MobileNetV2    & 69.3     & 69.4      & Heavy  & 67.2    & 70.5       & Heavy  & 68.7     & 72.8       & Heavy  \\
			SqueezeNet     & 37.7     & 67.5      & Medium & 29.0    & 29.0       & Medium & 26.7     & 68.6       & Medium \\ \hline
			& \multicolumn{3}{c}{Nexus S}   & \multicolumn{3}{c}{Pixel 2}  & \multicolumn{3}{c}{Pixel 3 XL} \\ \hline
			MobileNetV2    & 67.7     & 88.8       & Heavy  & 66.2    & 69.4       & Heavy  & 63.6     & 76.5       & Heavy  \\
			SqueezeNet     & 32.7     & 64.4       & Medium & 28.8    & 70.1       & Medium & 25.8     & 66.1       & Medium \\ \hline
		\end{tabular}}
	\end{center}
	\caption{Comparison of resource consumption of the two on-device MS networks deployed to the six Android based mobile devices. }\label{tab:resource}\vspace{-8mm}
\end{table*}

Duchesne et al. \cite{duchesne2020tracking} also report a high accuracy (0.827) of predicting the ``Worse" category based on the feature extracted from a single CXR with their highly imbalanced testing dataset, which contains over 84.6\% samples labeled as ``Worse". However, the reported accuracy is lower than a simple baseline: a dummy classifier that always predicts the most frequent label ``Worse" would yield a higher accuracy of 0.846. To make a comparison, we reimplement their model \cite{duchesne2020tracking} on our more balanced dataset and record a result of 0.600, which implies that using the feature from a single CXR may not be sufficient to predict radiological trajectory. On the other hand, by using feature concatenation from a series of longitudinal CXR images, our model demonstrates better and more reliable performance (0.800).

In order to systematically evaluate the performance of the MS networks under the different decision thresholds, we again use the AUROC value to assess how capable the model is in discriminating ``Worse" cases from ``Improved" cases and ``Stable" cases. As shown in Fig. \ref{fig:ROC}, MobileNetV2 shows a close performance compared to the RF network (DenseNet-121). It is important to note that MobileNetV2 networks can achieve a high AUROC value of 0.883 enabling it to identify ``Worse" cases from ``Improved" cases and show a significant potential of on-device follow-up using CXR images.

\section{Performance Evaluation on Mobile Devices}

For on-device COVID-19 patient triage and follow-up with resource constraints, resource consumption is also an important consideration for performance evaluation in addition to accuracy. In order to systematically  assess the performance of our COVID-19 on-device app, we select six mobile systems released following a chronic order, i.e., Nexus One / Nexus S (low-end); Pixel/ Pixel 2 (mid-range) and Pixel 2 XL/ Pixel 3 XL (high-end). Using the Pytorch Mobile framework, we deploy the three MS networks to the six Android based mobile systems and compare the resource consumption with regard to CPU, memory and energy usages. Fig. \ref{fig:android} describes a workflow to build an Android app based on the MS networks for on-device patient triage and follow-up.  

In Table \ref{tab:resource}, it is clear that the MobileNetV2 based COVID-19 app is resource-hungry, demonstrated by much higher resource consumption than SqueezeNet. Thus, the high accuracy achieved by MobileNetV2 is at the cost of high resource consumption. Within each app, we observe a downward trend in resource consumption following the chronic order, reflecting a continuous improvement of mobile device hardware. Overall, MobileNetV2 based COVID-19 apps are more suitable for high-performing mobile devices due to the high accuracy achieved with a higher resource consumption. On the other hand, SqueezeNet is more suitable for low-end mobile devices with both lower accuracy and resource consumption.

\section{Conclusions}
The classical two-player knowledge distillation framework \cite{hinton2015distilling} has been widely used to train a compact network that is hardware-friendly with ample applications \cite{dhar2019device}. In the related task of on-device natural image classification, the teacher network is pre-trained with ImageNet and distill the knowledge to a lightweight student network (e.g., MobileNetV2). This two-player framework, although seemingly successful, can be problematic for on-device medical imaging based COVID-19 case screening and radiological trajectory prediction described herein. The large gap between natural images and the medical images of a specific disease such as COVID-19 makes the knowledge distillation less effective as it is supposed to be. The small number of labeled COVID-19 images for training further aggravates the situation. 

In our three-player KTD framework, knowledge transfer from the AP network to the RF network can be viewed as a more effective regularization as they are built on the same network architecture, which in turn, make the knowledge distillation more effective since the RF network and MS network share the same training set. Different from what has extensively investigated focusing on the impact of distillation strength and temperature, we uncover a pivotal role of employing novel loss functions in refining the quality of knowledge to be distilled. Hence our three-player framework provides a more effective way to train the compact on-device model using a smaller dataset while preserving performance. 

From a more broad perspective, the three-player KTD framework is generally applicable to train other on-device medical imaging classification and segmentation apps for point-of-care screening of other human diseases such as lung \cite{wang2017chestx} and musculoskeletal \cite{rajpurkar2017mura} abnormalities.

\bibliographystyle{IEEEtran}  
\bibliography{references} 

\end{document}